\documentclass[a4paper,twoside,twocolumn]{article}
\pdfoutput=1

\usepackage{
url       %
, fancyhdr 
, lastpage 
, amsmath  
}

\usepackage[noindent, arraystretch, fullpage]{setlengths} 
\usepackage{
own         
}

\usepackage{ifpdf}  
\ifpdf
    \usepackage[pdftex]{graphicx}
    \usepackage[colorlinks,linkcolor=blue,citecolor=blue,urlcolor=blue]{hyperref}
    \DeclareGraphicsExtensions{.pdf}
\else
    \usepackage{graphicx}
    \usepackage[hypertex]{hyperref}  
\fi

\usepackage{todonotes} 

\graphicspath{{images/}}

\newcommand*{\metaauthori}{Bob Briscoe}
\newcommand*{\metaauthorii}{Koen De Schepper}
\newcommand*{\metashorttitle}{Scaling TCP's cwnd below 2\,MSS}
\newcommand*{\metatitle}{Scaling TCP's Congestion Window for Small Round Trip Times}
\newcommand*{\metano}{BT Technical Report; TR-TUB8-2015-002}
\newcommand*{\metakeywords}{Data Communications, Networks, Internet,
Control, Performance, Latency, Responsiveness, Dynamics, Algorithm, Standards}

\newcommand*{\metamaildi}{@bobbriscoe.net}
\newcommand*{\metamaili}{\href{mailto:ietf}{ietf}}
\newcommand*{\metamaildii}{@alcatel-lucent.com}
\newcommand*{\metamailii}{\href{mailto:koen.de\_schepper}{koen.de\_schepper}}
\newcommand*{\metaaddressi}{At the time of writing, Bob Briscoe was with BT Research \& Technology, Ipswich, UK}
\newcommand*{\metaaddressii}{Alcatel-Lucent, Antwerp, Belgium}

\newcommand*{\metaversion}{00E}
\newcommand*{\metadate}{10 Jun 2015}

\hypersetup{                       
     pdfauthor = {\metaauthori \& \metaauthorii},
     pdftitle = {\metashorttitle},
     pdfsubject = {},
     pdfkeywords = {\metakeywords}
}%

\title{\metatitle}%
\author{\metaauthori%
\thanks{\metamaili\metamaildi, %
\metaaddressi}%
\ %
\and \metaauthorii%
\thanks{\metamailii\metamaildii,%
\metaaddressii}%
}
\date{\metadate}%

\fancypagestyle{first}{%
\fancyhead[LO,RE]{}%
\fancyhead[LE,RO]{}%
\fancyhead[C]{}%
}%

\begin{document}
\bibliographystyle{alpha}%


\maketitle%
\thispagestyle{first}

\begin{abstract}
{\small\noindent%

This memo explains that deploying active queue management (AQM) to
counter bufferbloat will not prevent TCP from overriding the AQM and
building large queues in a range of not uncommon scenarios. This is 
a brief paper study to explain this effect which was observed in
a number of low latency testbed experiments.

To keep its queue short, an AQM drops (or marks) packets to make the
TCP flow(s) traversing it reduce their packet rate. Nearly all TCP
implementations will not run at less than two packets per round trip
time (RTT). 2pkt / RTT need not imply low bit-rate if the RTT is small.
For instance, it represents 2Mb/s over a 6ms round trip. When a few TCP
flows share a link, in certain scenarios, including regular broadband
and data centres, no matter how much the AQM signals to the flows to
keep the queue short, they will not obey, because it is impossible for
them to run below this floor. The memo proposes the necessary
modification to the TCP standard.
}      
\end{abstract}
\section{The Problem}\label{sec:submssw_Problem}

The capacity-seeking (aka.\ greedy) behaviour of TCP and its
derivatives has led to the need for active queue management (AQM) which
starts to drop packets as the queue grows, even when it is still quite
short. Then the queue stays short, and the rest of the buffer remains
available to absorb bursts.

Keeping down queuing delay, obviously drives down the round trip time
(RTT). For a certain number of flows sharing a link, the packet rate of
each will stay the same if the RTT reduces. But a lower RTT means less
packets \emph{per RTT}. Unfortunately, nearly all TCP implementations
cannot operate at less than two packets per RTT (the
standard~\cite{IETF_RFC5681:TCP_algorithms} prohibits it).

\begin{figure}[h]
  \centering
  \includegraphics[width=1.05\linewidth]{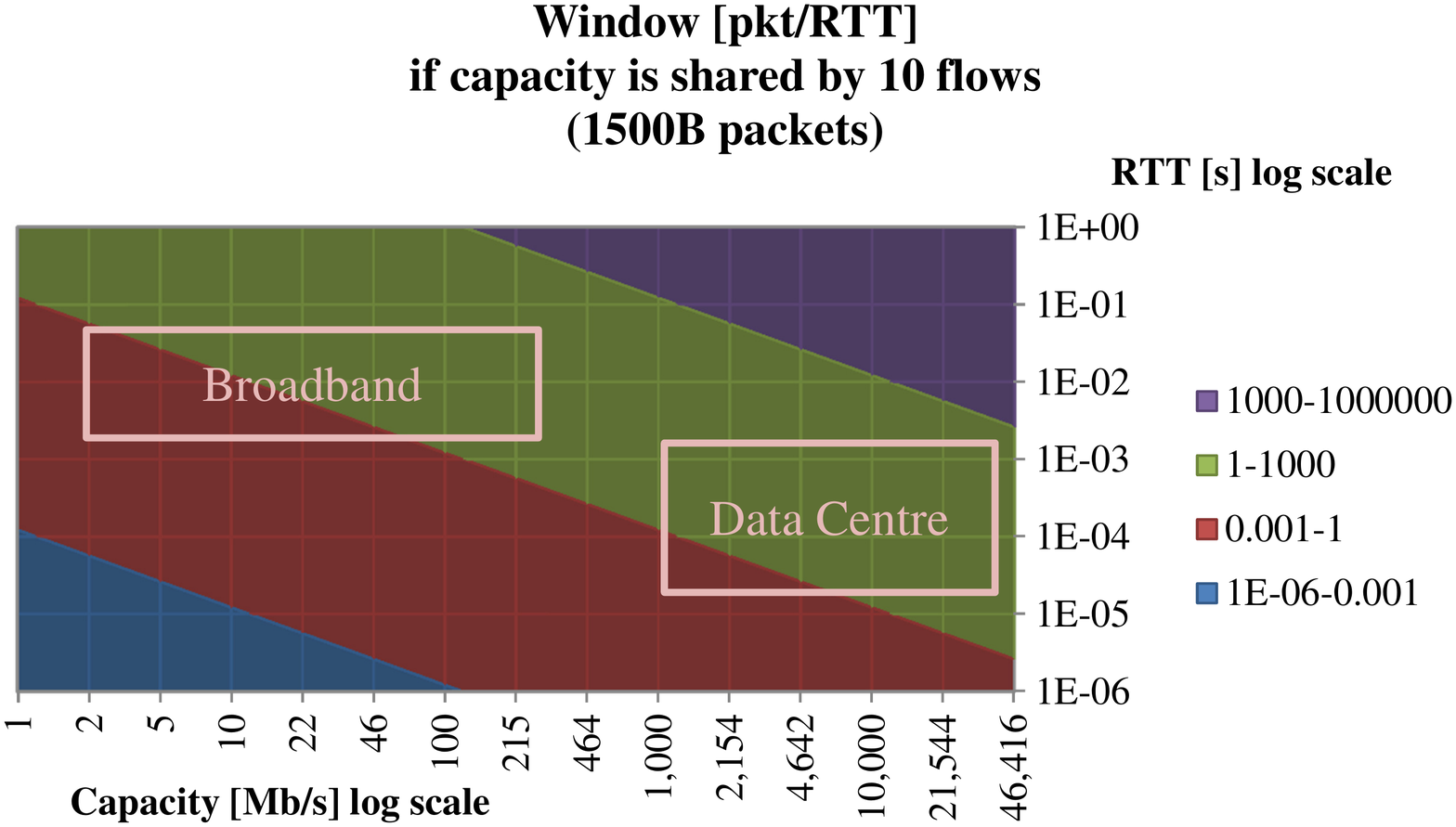}\\
  \caption{Window Size at Various Scales}\label{fig:submssw_scenarios}
  {\small{}Only the middle diagonal is significant for this study,
  representing a window of 1 MSS. The same diagonal would represent 2 MSS for 5 flows.}
\end{figure}

How common are these circumstances? Imagine a quite unremarkable
scenario in a residential broadband setting where 12 equal flows all
share a 40\,Mb/s link with an Ethernet frame size of 1518\,B, so each
sends at \(40/12 = 3.3\)\,Mb/s. That's not so slow. But if an AQM
attempts to keep their round trip time down to \(R=6\,ms\), they would
have to run at \(40M/(12
* 1518 * 8) * 6m = 1.64\)\,pkt/RTT. They cannot and will not run at less than 
2 pkt/RTT.

A scenario with a shorter round trip time, a slower link, more flows or
larger packets would require even less packets per round trip. In the
developing world sub-packet windows are much more
common~\cite{Chen11:TCP_sub-packet}. Nonetheless, taking an example
scenario of 10 flows sharing the bottleneck,
\autoref{fig:submssw_scenarios} illustrates how the developed world is
definitely not immune to the problem. The need for sub-single-packet
windows is probably not at all unusual in both broadband and data
centre scenarios. However, the prevalence of these scenarios in practice
will need to be established. For instance, in data centres, the configured 
maximum packet size is often larger (e.g.\ 9KB).

TCP controls its rate using a window mechanism, where the window, \(W\)
is the number of segments per round trip. The mechanism cannot work for
a window of less than two segments, and TCP's standard congestion
avoidance algorithm~\cite{IETF_RFC5681:TCP_algorithms} stipulates a
minimum window of 2*SMSS, where SMSS is the sender's maximum segment
size (usually 1460\,B). The 2 is intended to interwork with the common
delayed ACK mechanism that defers an ACK until a second segment has
arrived or the timer has expired (default 40\,ms in Linux).

Once TCP's window is at this minimum, TCP no longer slows down, no
matter how much congestion signalling the AQM emits. TCP effectively
ignores the increasingly insistent drops (or ECN marks) from the AQM.
Inside the algorithm it halves the congestion window, but then rounds
it back up to the minimum of two. For non-ECN flows, this will drive
the AQM to make the queue longer, which in turn will drop more packets.
So the flows will shuffle between periods waiting for timeouts and
periods going faster than average while others wait for timeouts (see
\cite{Morris97:TCP_many_flows}), but there will always be a longer
queue. ECN flows will just keep making the queue longer until the RTT
is big enough. In the following, where we don't need to distinguish ECN
and non-ECN, the term `signals' will be used for either drops or ECN
marks.

As long as TCP effectively ignores congestion signals, queuing delay
increases, the AQM emits even more signals, TCP's rounding-up
effectively ignores them, the queuing delay increases, and so on. TCP
is designed to reduce its window, not only when congestion signals
increase, but also as RTT increases. So the queue will eventually
stabilise at some larger size than the AQM would have liked (assuming
there is sufficient buffer above the AQM's target queuing delay).
Balance will be reached when all the flows are sending TCP's minimum
number of segments per round trip. Because, above that, all the TCPs
will reduce their window in response to any additional signals, but
below that they won't.

So, that's good isn't it?

No. The flows are indeed sharing the link (without any losses in the
case of ECN), they are clocking out packets twice every round trip and
everything is stable. But to achieve this they have overridden the AQM
to build a standing queue. A better outcome would be for all the TCPs
to send out packets less often than twice per round trip, which would
keep the queue at the level intended by the AQM.

Note that this problem is not the same as TCP's ``Silly Window
Syndrome''. Both problems do concern a sub-SMSS window, but the present
problem concerns the congestion window, not the flow control window.

Until now, the networkign community thought that AQM was the solution 
to the queuing delay caused by TCP's capacity-seeking behaviour. However, 
in these scenarios TCP still trump AQM.

\section{A Sub-MSS Window Mechanism for TCP}\label{sec:submssw_submssw}

No amount of AQM twiddling can fix this. The solution has to fix TCP.

In the following we consider how TCP might be modified to work internally 
with a fractional window instead of rounding it up to \(2*M\),  where we use 
the symbol \(M\) for SMSS.

The window mechanism is fundamentally a way to send \(W\)
bytes\footnote{Working in units of bytes not packets is necessary for
what follows} every RTT, \(R\).

We want to extend the window mechanism if \(W<2M\) to send a packet of
\(M\) bytes every \(M/W\) round trips.

It is always wrong to send smaller packets more often, because the
constraint may be packet processing, not bits. So we will not send
smaller than maximum sized segments unless the send queue is
insufficient. Therefore, more generally, if \(W<2s\) we want to send a
packet of size \(s\) bytes every \(s/W\) round trips, where
\(s=\mathrm{min}(M,\mathit{snd\_q})\), and where \(\mathit{snd\_q}\) is
the amount of outstanding data waiting to be sent.

\begin{figure}[h]
  \centering
  \includegraphics[width=0.4\linewidth]{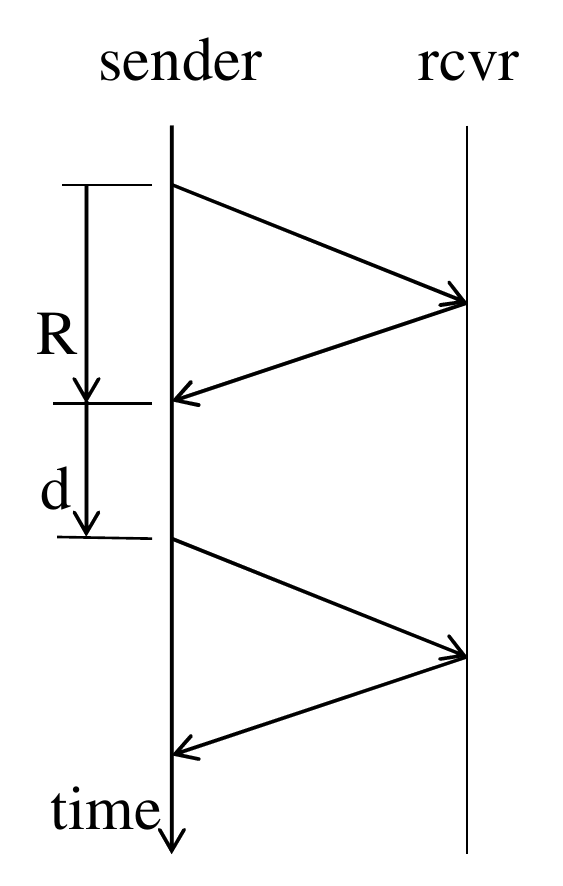}\\
  \caption{TCP's Segment Timing for \(W<1\)}\label{fig:submssw_approach}
\end{figure}

Normally, TCP holds back from sending if \(W<s\). As illustrated in
\autoref{fig:submssw_approach}, we need to modify this so that,
following receipt of an ACK, if \(W<s\) TCP waits for time \(d\),
where:
\begin{equation*}
   d + R = \frac{sR}{W}
\end{equation*}
\begin{equation}
   d = \left(\frac{s}{W}-1\right)R.
   \label{eqn:submssw_d}
\end{equation}
Ironically, the sender has to insert a delay between each packet to
avoid delay in the queue. Indeed, it is the same amount of delay. But
it is better to localize the delay at each sender, not remote in the
network, otherwise:
\begin{description}
  \item[Mutual interest:] There is a strong possibility that the
      remote queue is shared by other flows, some of which are likely
      to be interactive;
  \item[Self-interest:] Many modern apps (e.g.\ HTTP/2, SPDY and
      interactive video) can adapt what they send and how much they
      send based on the local send queue, whereas it takes a round
      trip to know the status of a remote network queue;
\end{description}

The way to determine how best to implement the above is to follow the
approach of TCP Laminar~\cite{Mathis12:Laminar_TCP_ID}. That is to
focusing solely on the part that keeps TCP `clocking' at a constant
rate, and treat the ups and downs that adjust that rate (congestion
control, flow control) and regulate ACKs (delayed ACK) as separate 
(interactions will need to be considered, but only later). 

A TCP sender's basic window clocking machinery normally works as follows: when
TCP sends \(s\) bytes, it decrements \(W\) by \(s\) and when the sender
receives an acknowledgement for \(s\) bytes one RTT later, it
increments \(W\) by \(s\).

This needs to be modified as follows: after 
TCP's congestion response
following receipt of an acknowledgement, if \(W<s\), TCP must wait
\((s/W - 1)R\), which then entitles it to send a packet of size \(s\)
and decrement \(W\) by \(s\). This makes \(W\) negative, which is
conceptually OK, but would require major change across TCP. See later
for possible alternative ways to implement this. However, for now bear
with the conceptual device of a negative window. Then the behaviour (as
opposed to the implementation) of the rest of TCP should not need to
change.

Specifically, unless other parts of TCP have intervened in the
meantime, TCP will receive the acknowledgement for \(s\) bytes and
increment the window by \(s\). \(W\) should now be positive again, but
still insufficient to send a packet of size \(s\). So TCP must again
wait \((s/W - 1)R\).

The congestion control parts of TCP's machinery will be independently
increasing \(W\) every time an ACK is received or reducing it every
NACK. If congestion of the link is relieved, as each packet is ACKed,
TCP will increase \(W\) and consequently reduce the wait \(d\) between
packets. As \(W\rightarrow s\) the wait \(d\rightarrow 0\). Once
\(W\geq s\) no wait will be necessary. If instead the link remains
congested, every time a NACK is received, \(W\) will reduce (e.g.
halve), and the wait between packets will increase appropriately (e.g.\
approximately double).

The normal congestion responses of a set of TCP's with the above
modification should work properly with the AQM at a bottleneck. They
would pace the segments at less than one per round trip if necessary.
Then they will balance with the AQM at its intended queuing delay,
rather than bloating the queue just so they can all run at 2 segments
per RTT.

This mechanism should be able to replace TCP's exponential back-off, as
a more justifiable way to keep a congested link just busy enough during
congestion. Every time TCP's retransmission timer expires, it will
halve \(W\), thus doubling the wait \(d\) before sending the next
retransmission. With no response, \(W\) will get exponentially smaller
and \(d\) exponentially larger. But as soon as there is one ACK, the
window will grow so that a data packet (or probably a retransmission)
can be sent.

\section{Potential Issues}\label{sec:submssw_issues}

Even if \(d\) is large relative to \(R\), TCP will have to use the last
estimate of \(R\) because it will have no better way to estimate \(R\)
given all activity will have stopped.

Modifying TCP implementations is unlikely to be straightforward.
Integer arithmetic will need to be developed for
\autoref{eqn:submssw_d}. A sub-MSS window has been implemented in Linux
before, in TCP Nice, but the code is now quite old. Also many parts of
TCP are likely to have to be changed to implement the concept of a
negative window. It would not be appropriate to store the amount of
negativity in a separate variable in order to limit side-effects,
because the whole point of the window variable is to communicate
side-effects to all the different parts of the code.

If the type of the window variables were changed from unsigned to
signed, this would lose half of the maximum window size. A better
alternative might be to increment up the meaning of cwnd by one SMSS,
and appropriately change all the places where it is compared with zero
or other constants, such as SMSS. However, for widely understood code
like TCP, such a change could cause implementers to get very confused.

TCP's delayed ACK mechanism causes only every \(n\) (default 2)
segments arriving at the receiver to elicit an ACK, unless more than
the delayed ACK timer (default 40\,ms in Linux) elapses between
packets. Assuming the receiver delays ACKs, the above sub-MSS mechanism
will result in all the segments being sent at the correct average rate,
but in pairs. That is OK, but not ideal. The idea in
AccECN~\cite{Briscoe14d:accecn_ID} where the sender can ask the
receiver to turn off delayed ACKs would be nice.

A priority would be to support a sub-MSS window in data centre TCP
(DCTCP~\cite{Alizadeh10:DCTCP}) before it is deployed over the public
Internet as proposed~\cite{Kuehlewind14a:DCTCP_Internet,
deSchepper15a:CoupledAQM}. This is because DCTCP can maintain an
extremely shallow queue, so it will more often need a window below one
to support this (we uncovered the present problem while testing DCTCP
over a broadband access---more than a certain number of flows suddenly
started the queue growing).

\section{Related Work}\label{sec:submssw_Related}

Morris~\cite{Morris97:TCP_many_flows} found that much of the loss in
the Internet in 1997 was due to many TCP flows at bottlenecks, causing
an average window of less than one segment, which actually appears as a
shuffling between some flows waiting for time-outs while others consume
much more than the equal share. As a sign of the times, one proposed
solution was to add more buffer space although it was recognised a more
fundamental solution was really needed, for which RED was suggested,
although as this memo points out, that would not have helped.

TCP Nice~\cite{Venkata02:TCP_Nice} is intended for background
transport. It is a modification to TCP Vegas in Linux to make it more
sensitive to congestion and includes support for less than one segment
in the congestion window. When the window is below 2 segments, it
switches into a new mode and sends a packet every \(1/W\) round trips,
which might be an easier approach for implementation, but it might not
behave as continuously across the range of window values as the above
proposal. TCP Nice arbitrarily limits the minimum window size to
\(1/48\). The paper also reports on a simulation of Nice's interaction
with the RED AQM.

Chen et al.~\cite{Chen11:TCP_sub-packet} investigates the behaviour of
TCP when the path can only support a window of less than one segment,
primarily interested in oversubscribed low capacity links in the
developing world.

Komnios et al.~\cite{Komnios14:LEDBAT_sub-packet} find that LEDBAT
performs better than TCP in the regime with a window of less than one
segment, but when there is a mix of flows on the link, LEDBAT switches
into its TCP mode, so it needs to be more sophisticated when it can do
better by remaining in LEDBAT mode.

\section*{Acknowledgements}\label{sec:submssw_Ack}

Thanks to Michael Welzl, John Heffner, Joe Touch and Matt Mathis for explaining
why the minimum ssthresh in the TCP spec is \(2*\mathrm{SMSS}\), and to
Michael Welzl and Anna Br\"unstrom for helping with the Related Work.

The authors were part-funded by the European Community under its Seventh
Framework Programme through the Reducing Internet Transport Latency
(RITE) project (ICT-317700). The views expressed are solely those of
the authors.

\addcontentsline{toc}{section}{References}

{\footnotesize%
\bibliography{bob}}

\newcommand{\etalchar}[1]{$^{#1}$}
\begin{thebibliography}{DSTBB14}

\bibitem[AGM{\etalchar{+}}10]{Alizadeh10:DCTCP}
Mohammad Alizadeh, Albert Greenberg, David~A. Maltz, Jitu Padhye, Parveen
  Patel, Balaji Prabhakar, Sudipta Sengupta, and Murari Sridharan.
\newblock {Data Center TCP (DCTCP)}{}.
\newblock {\em Proc. ACM SIGCOMM'10, Computer Communication Review},
  40(4):63--74, October 2010.

\bibitem[APB09]{IETF_RFC5681:TCP_algorithms}
M.~Allman, V.~Paxson, and E.~Blanton.
\newblock {TCP Congestion Control}{}.
\newblock Request for Comments 5681, RFC Editor, September 2009.

\bibitem[BKS18]{Briscoe14d:accecn_ID}
Bob Briscoe, Mirja K\"uhlewind, and Richard Scheffenegger.
\newblock {More Accurate ECN Feedback in TCP}{}.
\newblock Internet Draft draft-ietf-tcpm-accurate-ecn-07, Internet Engineering
  Task Force, July 2018.
\newblock (Work in Progress).

\bibitem[CISF11]{Chen11:TCP_sub-packet}
Jay Chen, Janardhan Iyengar, Lakshminarayanan Subramanian, and Bryan Ford.
\newblock {TCP Behavior in Sub-Packet Regimes}{}.
\newblock In {\em Proc. SIGMETRICS'11}, pages 157--158. ACM, June 2011.

\bibitem[DSTBB14]{deSchepper15a:CoupledAQM}
Koen De~Schepper, Ing-Jyh Tsang, Olga Bondarenko, and Bob Briscoe.
\newblock {Data Center to the Home}{}.
\newblock Presentation in IETF Proceedings, URL:
  \url{http://www.ietf.org/proceedings/92/slides/slides-92-iccrg-5.pdf}, March
  2014.

\bibitem[KSC14]{Komnios14:LEDBAT_sub-packet}
I.~Komnios, A.~Sathiaseelan, and J.~Crowcroft.
\newblock Ledbat performance in sub-packet regimes.
\newblock In {\em Wireless On-demand Network Systems and Services (WONS), 2014
  11th Annual Conference on}, pages 154--161, April 2014.

\bibitem[KWEB14]{Kuehlewind14a:DCTCP_Internet}
M.Mirja K\"uhlewind, David~P. Wagner, Juan Manuel~Reyes Espinosa, and Bob
  Briscoe.
\newblock {Using Data Center TCP (DCTCP) in the Internet}{}.
\newblock In {\em Proc. Third IEEE Globecom Workshop on Telecommunications
  Standards: From Research to Standards}, pages 583--588, December 2014.

\bibitem[Mat12]{Mathis12:Laminar_TCP_ID}
Matt Mathis.
\newblock {Laminar TCP and the case for refactoring TCP congestion control}{}.
\newblock Internet Draft draft-mathis-tcpm-tcp-laminar-01, Internet Engineering
  Task Force, July 2012.
\newblock (Work in progress).

\bibitem[Mor97]{Morris97:TCP_many_flows}
R.~Morris.
\newblock {TCP Behavior with Many Flows}{}.
\newblock In {\em Proceedings of the 1997 International Conference on Network
  Protocols (ICNP '97)}, ICNP '97, pages 205--, Washington, DC, USA, 1997. IEEE
  Computer Society.

\bibitem[VKD02]{Venkata02:TCP_Nice}
Arun Venkataramani, Ravi Kokku, and Mike Dahlin.
\newblock {TCP Nice: A Mechanism for Background Transfers}{}.
\newblock {\em SIGOPS Oper. Syst. Rev.}, 36(SI):329--343, December 2002.

\end{thebibliography}


\onecolumn%
\addcontentsline{toc}{part}{Document history}
\section*{Document history}

\begin{tabular}{|c|c|c|p{3.5in}|}
 \hline
Version &Date &Author &Details of change \\
 \hline\hline
00A          &15 May 2015 &Bob Briscoe &First Draft \\\hline%
00B          &15 May 2015 &Bob Briscoe &Added Abstract, Scenarios and Related Work\\\hline%
00C          &15 May 2015 &Bob Briscoe &Included design approach and implementation issues.\\\hline%
00D          &\metadate   &Bob Briscoe &Removed mistaken idea that \(1*\mathrm{SMSS}\) would be easier than \(<1*\mathrm{SMSS}\).\\\hline%
\metaversion &\metadate   &Bob Briscoe &Added KDS as author. Corrected inconsistencies.\\\hline%
\hline%
\end{tabular}

\end{document}